\documentclass[a4paper]{jpconf}
\usepackage[utf8]{inputenc}
\usepackage{amsmath}
\usepackage{amsfonts}
\usepackage{amssymb}
\newcommand{\nn}{\nonumber}
\begin{document}
\title{Magnetic monopoles and symmetries in noncommutative space}

\author{Samuel Kov\'a\v{c}ik}

\address{Dublin Institute for Advanced Studies, 10 Burlington Road, Dublin 4, Ireland}

\ead{skovacik@stp.dias.ie}

\author{Peter Prešnajder}

\address{Faculty of Mathematics, Physics and Informatics, Comenius University Bratislava, Mlynsk\'a dolina, Bratislava, 842 48, Slovakia}

\ead{presnajder@fmph.uniba.sk}
\begin{abstract}
In this paper, we review the progress in the analysis of magnetic monopoles as generalized states in quantum mechanics. We show that the considered model contains rich algebraic structure that generates symmetries which have been utilized in different physical contexts. Even though are we focused on quantum mechanics in noncommutative space $\textbf{R}_\lambda^3$, the results can be reconstructed in ordinary quantum mechanics in $\textbf{R}^3$ as well.\\
\textbf{Keywords:} Magnetic monopoles, quantum mechanics, noncommutative space \\
\\
Ref. number: DIAS-STP-17-09 \\
Prepared for conference proceedings of \textit{International Conference on Integrable Systems and Quantum symmetries 2017}
\end{abstract}

\section{Introduction}
It is generally believed that the Planck units describe important scales where aspects of both the quantum world and the curved space-time being relevant. It is perfectly safe to assume that the laws of nature looked different when the Universe had the Planck temperature. Similarly, we can expect the space to have a nontrivial structure on the scale of the Planck length. Quantum corrections to gravity cannot be neglected and vice versa. 

The existence of a fundamental length scale is predicted by most of the candidate theories of quantum gravity. That it shall be no surprise can be argued from already well-known facts. For example, the original argument of Snyder \cite{snyder} states that if you want to distinguish two space points separated by the Planck length you need to use a photon of a similar wavelength. However, such a photon would be hidden under its event horizon – a black hole would form and no information could be obtained. It is natural to abandon the notion of exact space(time) points in a similar manner as we disregarded the classical notion of position and momentum of a particle when formulating quantum mechanics. 

Correspondingly to the phase space in quantum mechanics, a space whose coordinates cannot be distinguished is described by a general noncommutativity (NC) relation 
\begin{equation}
[x_i,x_j] = i\lambda \Theta_{ij} (x),
\end{equation}
where $\Theta_{ij}$ is antisymmetric in indices $i,j$ and $\lambda$ describes the scale of noncommutavity, which is often assumed to be of the order of the Planck length. There are many possible choices for the range (and structure) of indices $i,j$ and for $\Theta$. A popular choice is to take $i,j=1,2$ and constant $\Theta$, the so called Groenewold-Moyal plane \cite{groen, moyal}. However, we will be interested in the case of $\Theta_{ij}(x) = 2 \varepsilon_{ijk} x_k$, where the indices $i,j,k$ take values of $1,2,3$ and $\varepsilon$ is the Levi-Civita symbol. Therefore, our starting point is the following NC relation
\begin{equation} \label{NCrel}
[x_i, x_j] = 2 i \lambda \varepsilon_{ijk} x_k.
\end{equation}

This choice describes a 3D rotationally invariant space $\textbf{R}_\lambda^3$, which can be viewed as a sequence of concentric fuzzy spheres $\textbf{S}^2_\lambda$ of different radii. There are many ways of constructing NC coordinates, we will utilize the construction using (two sets of) creation and annihilation (c/a) operators satisfying the usual relations
\begin{equation}\label{aux}
[a_\alpha,a^+_\beta]=\delta_{\alpha\beta },\ \
[a_\alpha,a_\beta]=[a^+_\alpha, a^+_\beta]=0
\end{equation} 
and acting in an auxiliary Fock space $\mathcal{F}$ spanned on normalized states 
\begin{equation}
|n_1,n_2\rangle= \frac{(a^+_1)^{n_1}\,(a^+_2)^{n_2}}{
\sqrt{n_1!\,n_2!}}\ |0\rangle .
\end{equation}

The NC coordinates are defined as 
\begin{equation} \label{ncX}
x^i = \lambda \sigma^i_{\alpha \beta} a^+_\alpha a_\beta.
\end{equation}
Each part of this expressions serves its own purpose: the c/a operators bring in the NC property, the Pauli matrices $\sigma^i$ generate the $su(2)$ structure with coefficients $f_{ijk} = \varepsilon_{ijk}$ and $\lambda$ carries the length scale.  We also define the radial distance coordinate as
\begin{equation}
r = \lambda \left(a^+_\alpha a_\alpha +1 \right),
\end{equation}
where the $+1$ was added to ensure that $r^2 - x_i^2 = O(\lambda^2)$ (so there is no term linear in $\lambda$). 

This construction of 3D rotational invariant space $\textbf{R}_\lambda^3$ was developed in \cite{Jabbari} and explored by the present authors and their collaborator in \cite{GP1,GP2, vel, LRL, plane, mbh, mm}. 

\section{Hilbert space $\mathcal{H}$ and subspace $\mathcal{H}_0$}
We can now formulate quantum mechanics (QM) in $\textbf{R}_\lambda^3$. This theory, which is being referred to as NC QM, will differ from the ordinary QM unless we send $\lambda \rightarrow 0$, which is often being referred to as taking the commutative limit.

A natural choice for the Hilbert space $\mathcal{H}$ is the set of functions of the c/a operators \eqref{aux}. Most of the previous work was focused on states containing an equal number of creation as annihilation operators. Restricting only on this subspace, which is denoted $\mathcal{H}_0$, makes a perfect sense. Note that in the definition of NC coordinates \eqref{ncX} each of the coordinates contains exactly one creation and one annihilation operator. Therefore any state of the form $\Psi(x)$ should be in $\mathcal{H}_0$ (and actually nothing else is). 

The Hilbert space $\mathcal{H}$ contains states of the form $\Psi\ =\ \Psi(a^+,a)$ and is equipped with a norm

\begin{equation} \label{norm}
 \| \Psi \|^2\  =\ 4\pi\,
\lambda^2\,\mbox{Tr}[\Psi^\dagger \hat{r} \Psi] ,
\end{equation}
where $\hat{r} \Psi = \frac{1}{2} \left( r \Psi + \Psi r \right)$. Let us now recall the definition of some important operators. Because of the $su(2)$ structure appearing in \eqref{NCrel} it is not hard to define the generators of rotations $\hat{L}_i$ satisfying the $su(2)$ relations, they follow as
\begin{eqnarray} \label{gene}
\hat{L}_i \Psi &=& \frac{1}{2 \lambda}[x_i, \Psi], \\ 
\hat{X}_i \Psi &=& \frac{1}{2} \{ x_i, \Psi \}, 
\end{eqnarray}
where in the second line we have defined the symmetrical action of the coordinate operator. In \cite{vel} it has been shown that these two can be combined into a single object $\hat{L}_{ij}$ generating $SO(4)$ rotations. 

The free Hamiltonian has been defined in \cite{GP1} as 
\begin{equation}
\hat{H}_0 \Psi =  \frac{1}{2m \lambda \hat{r}}\, [a^+_\alpha , [a_\alpha , \Psi]],
\end{equation}
where it was used to define the hydrogen atom problem, which was solved in the same paper by directly solving the corresponding Schr\"odinger equation and in \cite{GP1,GP2} algebraically by the means of Laplace-Runge-Lenz vector. 

This operator was also used in \cite{vel} to define the velocity operator (by the Heisenberg relation $\hat{V}_i = i [\hat{H}_0, \hat{X}_i]$) and $\hat{V}_4$ as
\begin{equation}
\hat{V}_i \Psi= \frac{i}{2\hat{r}}\sigma^i_{\alpha \beta} \left(a^+_\alpha \Psi a_\beta - a_\beta \Psi a^+_\alpha \right), \hat{V}_4 = \frac{1}{2 \hat{r}}  \left(a^+_\alpha \Psi a_\alpha+ a_\alpha \Psi a^+_\alpha \right) = \lambda^{-1} \left(1- \lambda^2\hat{H}_0 \right) \Psi. 
\end{equation}

It was shown in the same paper that the four-vector $\hat{V}_a$ transforms under the $SO(4)$ rotations generated by \eqref{gene} and squares to a constant $\hat{V}_a^2 = \lambda^2$, which demonstrates the UV regularization of the theory. This result holds only for states in $\mathcal{H}_0$, but it has been generalized to $\mathcal{H}_\kappa$ (which is defined in the next section) in \cite{mm}.

Another important result holding only for $\mathcal{H}_0$ is that the components of $\hat{V}_i$ do commute with each other, even though the coordinate operators do not. 

The eigenspectrum of this operator has been found in \cite{plane}, the eigenfunctions have the same form as in the ordinary QM (that is $e^{i \vec{k} \cdot \vec{r}}$), but the eigenvalues have a(n upper) cut-off $ |k| \rightarrow \lambda^{-1} \sin \lambda |k|$. 

It should be noted that all of the presented results were exact (and not only perturbative), which signalizes symmetries are at work (and perhaps not all of them have been revealed yet). A downside of the model is that some of the famous problems we would expect to solve easily have not been solved as of now – the linear harmonic oscillator being a prominent example. Perhaps, the relevant symmetry is yet to be found.

\section{Generalized Hilbert space $\mathcal{H}_\kappa$}
Let us now move our attention to a generalized class of states which was investigated in \cite{mm}. It contains functions containing an unequal number of creation and annihilation operators, with their difference held fixed and equal to $\kappa$
\begin{equation} \label{states}
\Psi_\kappa(e^{-i\gamma} a^+, e^{i \gamma} a) = e^{-i\gamma\kappa} \Psi_\kappa (a^+,a), \ \gamma \in  \textbf{R}, \  \kappa \in \textbf{Z}  ,
\end{equation}
the corresponding Hilbert space is denoted $\mathcal{H}_\kappa$. We can use the same physical operators as for $\mathcal{H}_0$, yet there is one crucial difference now. 
From the fact that $[r , x_i]=0$ it follows immediately that $[\Psi(x_i) , r]=0$, so when working in $\mathcal{H}_0$ multiplying the states with $r$ from the left and from the right is the same and the calculations can often be simplified. Now the left and the right multiplication differ
\begin{equation*}
\hat{r}_L - \hat{r}_R = \lambda \kappa \neq 0
\end{equation*}
and one has to go through all of the calculations and correct them correspondingly. 

Let us now write down directly the results. For the sake of the point we will be making shortly we gather the relations describing a QM system containing a magnetic monopole of a charge $\mu$ as derived in \cite{zwanziger} on the left-hand side and the relations holding for $\mathcal{H}_\kappa$ on the right-hand side

\begin{eqnarray} \label{NCmm}
\ \left[ \hat{x}_i , \hat{x}_j \right]  = 0  & \leftrightarrow & [ \hat{X}_i, \hat{X}_j ] = \lambda^2 \varepsilon_{ijk} \hat{L}_k,   \\  \nn
 \left[ \hat{x}_i, \hat{p}_j  \right]  =  i \delta_{ij} & \leftrightarrow & [\hat{X}_i,\hat{V}_j] = i \delta_{ij} \left(1-\lambda^2 \hat{H}_0\right), \\  \nn
\ \left[\hat{p}_i ,\hat{p}_j  \right]  = i \mu \varepsilon_{ijk} \frac{\hat{x}_k}{r^3}  & \leftrightarrow & \ \left[\hat{V}_i,\hat{V}_j \right]  = i\frac{-\kappa}{2} \varepsilon_{ijk} \frac{\hat{X}_k}{\hat{r}(\hat{r}^2-\lambda^2)}, \\ \nn
 \hat{C}_1 = -\mu q   & \leftrightarrow &  \hat{C}_1= \frac{\kappa}{2}q , \\ \nn
 \hat{C}_2 = q^2 + (\mu)^2(-2E) & \leftrightarrow & \ \hat{C}_2 = q^2 +
\left(\frac{\kappa}{2}\right)^2 (-2E+\lambda^2 E^2).
\end{eqnarray}
$\hat{C}_1, \hat{C}_2$ are the Casimir operators for the hydrogen atom problem, which can be used to directly derive the energy spectrum without needing to solve the corresponding Schr\"odinger equation. 

If we set $\lambda=0$ it is obvious that the relations coincide if we identify $\mu = - \frac{\kappa}{2}$. We have to be careful since $\mu$ has to satisfy the Dirac quantization condition \cite{dirac}. However, the identification is perfect since $\kappa$ describes the difference between the number of creation and annihilation operators it follows that $\frac{\kappa}{2} \in \textbf{Z}/2$, which is the same condition as has to be satisfied by $\mu$. It was therefore concluded in \cite{mm} that $\mathcal{H}_\kappa$ is the Hilbert space of monopoles states of strength $- \frac{\kappa}{2}$.  

Monopole states can in some scenarios cause the non-associativity of the algebra \cite{mmna1,mmna2}. This is however not the case here since it can be shown that

\begin{equation} \label{NA}
\varepsilon_{ijk} [\hat{V}_i,[\hat{V}_j, \hat{V}_k]] = 0.
\end{equation}
This results is obvious as the operator product in a Hilbert space is associative, but is not very straightforward to prove as the zero comes from a cancellation of two rather nontrivial terms. 
 
\section{Underlying algebraic structure}

The fact that the generalization of the results could be carried out in such a simple way suggests that the underlying algebraic structures remain (nearly) unaffected by the presence the monopole states. Let us now investigate this issue. 

The NC space was defined using the c/a operators \eqref{aux} and so was the Hilbert space $\mathcal{H}$ and operators on it. In general, the c/a operators can act on the states in $\mathcal{H}$ either from the left or from the right. Let us denote these actions as

\begin{eqnarray}
\hat{a}_\alpha \Psi = a_\alpha \Psi, && \hat{a}_\alpha^+ \Psi = a_\alpha ^+ \Psi , \\ \nn
\hat{b}_\alpha \Psi =  \Psi a_\alpha, && \hat{b}_\alpha^+ \Psi =  \Psi a_\alpha ^+ .\\ \nn
\end{eqnarray}

We now have 4 annihilation (and 4 creation) operators, two acting from the left and two from the right. Let us combine them into a single object
\begin{equation}
\hat{A} = (\hat{a}_1, \hat{a}_2, \hat{b}_1, \hat{b}_2)^T, \ \hat{A}^+ = (\hat{a}_1^+, \hat{a}_2^+, \hat{b}_1^+, \hat{b}_2^+)
\end{equation}
Since multiplying from the right reverse the order the sign of the commutators is different 

\begin{equation}
[\hat{a}_\alpha, \hat{a}_\beta^+] = - [ \hat{b}_\alpha,\hat{b}_\beta^+] = \delta_{\alpha \beta} 
\end{equation}
and therefore $[\hat{A}_a, \hat{A}^+_b] \neq \delta_{ab}$ where $a,b =1,...,4$. To achieve this we need to add a matrix $\Gamma =\sigma_3 \otimes \textbf{1}_2 $ flipping the corresponding sings. Then it holds that
\begin{equation}
[\hat{A}_a, \Gamma_{bc} \hat{A}_c^+ ] = \delta_{ab}.
\end{equation}

We can now introduce a set of $su(2,2)$ matrices $S_{AB}=-S_{BA}$, where $A,B=0,...,5$, as

\begin{eqnarray}
S_{ij} = \frac{1}{2}\varepsilon_{ijk} \left( \begin{array}{cc}
\sigma_k & 0 \\ 
0 & \sigma_k 
\end{array} \right), && S_{k4} = \frac{1}{2}\left( \begin{array}{cc}
\sigma_k & 0 \\ 
0 & \sigma_k 
\end{array} \right), \\ \nn
S_{k5} = \frac{1}{2}  \left( \begin{array}{cc}
0& \sigma_k  \\ 
-\sigma_k & 0
\end{array} \right), && S_{45} = \frac{1}{2}\left( \begin{array}{cc}
0 & i \\ 
i &0 
\end{array} \right) , \\ \nn
S_{0k} = \frac{1}{2}  \left( \begin{array}{cc}
0& i\sigma_k  \\ 
i\sigma_k & 0
\end{array} \right), && S_{04} = \frac{1}{2} \ \left( \begin{array}{cc}
0 & 1 \\ 
-1 &0 
\end{array} \right) , \\ \nn
S_{05} = \frac{1}{2}\left( \begin{array}{cc}
 1 & 0 \\ 
0& -1 
\end{array} \right)= \frac{1}{2} \ \Gamma &&
\end{eqnarray}
to define operators 
\begin{equation}
\hat{S}_{AB} = \hat{A}^+ \Gamma S_{AB} \hat{A}
\end{equation}
forming a representation of the same algebra. 

All of the essential objects in the considered model of NC QM can be (in hindsight) expressed using $\hat{S}_{AB}$, as is expressed in the following table: 

\begin{center}
\begin{tabular}{|c|c|c|c|c|c|c|c|}
\hline 
$ \hat{S}_{AB}$ & $ \hat{S}_{ij}$ & $\hat{S}_{k4}$ &$ \hat{S}_{05}$ & $\hat{S}_{k5}$ & $\hat{S}_{45}$ & $ \hat{S}_{0k}$ & $\hat{S}_{04}$ \\ 
\hline 
$ \sim $ & $\varepsilon_{ijk} \hat{L}_k$ & $\hat{X}_k$ &$ \hat{r}$ & $ \hat{r}\bar{V}_k$ & $ \hat{r}\bar{V}$ & $\hat{r}\hat{V}_k$ & $\hat{r}\hat{H}_0$ \\ 
\hline 
\end{tabular} 
\end{center}
where $\hat{V}$ is the dilatation operator and $\bar{V}_k \Psi = \frac{1}{2 \hat{r}} \sigma^i_{\alpha \beta} \left( a^+_\alpha \Psi a_\beta + a_\beta \Psi a^+_\alpha \right)$. Up to a numerical factor, $\bar{V}_k$ is proportional to $\hat{\zeta}_k$, which was used to write down the Laplace-Runge-Lenz vector in \cite{LRL}.

The operators $\hat{L}_k$ and $\hat{X}_k$ commute with the weight factor $\hat{r}$ in \eqref{norm} and consequently they are Hermitian with respect to this norm. On the other hand, the operators $\hat{V}_k$, $\bar{V}_k$ and $\hat{H}_0$ do not commute with it and so the factor $\hat{r}^{-1}$ has to be added to ensure this.

The crucial realization is that this structure is completely indifferent of $\kappa$. Its only presence is in the quadratic operator $\hat{\mathcal{L}} = \hat{a}^+_\alpha \hat{a}_\alpha - \hat{b}_\alpha \hat{b}^+_\alpha$ that counts the difference between the creation and annihilation operators and is therefore proportional to $\kappa$.  

It shall be also noted that the $su(2,2)$ is the least you can utilize within the model, not the most. For example adding a factor $f(\hat{r})$ to $\hat{S}_{AB}$ differs the considered commutation relations a lot. Let us demonstrate it on the following example. We take two of the elements of the $su(2,2)$ and commute them

\begin{equation}
[\hat{S}_{0i},\hat{S}_{0j}] \propto \hat{S}_{ij} .
\end{equation}
The result is, obviously, in the same algebra. However, if we add the factor $\hat{r}^{-1} \propto \hat{S}_{05}$ it holds that

\begin{equation}
\lambda^{-1}[\hat{r}^{-1}\hat{S}_{0i},\hat{r}^{-1}\hat{S}_{0j}] \propto \frac{\kappa}{\hat{r}\left(\hat{r}^2 - \lambda^2\right)}  \hat{S}_{k4},
\end{equation}
or in other words, the third equation in \eqref{NCmm}. Thus, the monopole magnetic field appears as a correction due to the factors $\hat{r}^{-1}$ in the commutator (22). An analogous effects is
present in all other commutators $ [ \hat{r}^{-1} \hat{S}_{i5}, \hat{r}^{-1}, \hat{S}_{j5} ], [ \hat{r}^{-1} \hat{S}_{0i}, \hat{r}^{-1}, \hat{S}_{j5} ], ... $ associated with all noncompact $so(4,1)$ generators $\hat{S}_{0i}, \hat{S}_{04}, \hat{S}_{i5}$ and $\hat{S}_{45}$. Various components has been utilized this way, for example in \cite{vel} to explore the Euclidean kinematic $E(4)$ symmetry or in \cite{LRL} to describe the $SO(4)$ and $SO(1,3)$ symmetries of the hydrogen atom. 

\section{Conclusions}

We have reviewed the progress in the study of NC QM, focusing mostly on the very natural appearance of magnetic monopole states. The novel results presented here is firstly the $su(2,2)$ symmetry, which appears as a building block of the theory and is indifferent of considering the monopole states and secondly the fact that neither the nonassociativity of the model is violated by it \eqref{NA}.

The results of this paper can be reconstructed in QM in ordinary (commutative) space. One first needs to formulate the theory in $\textbf{C}^2$ instead of $\textbf{R}^3$ and replace the c/a operators with (complex) coordinates $\sqrt{\lambda} a_\alpha \rightarrow z_\alpha, \ \sqrt{\lambda}a^+_\alpha \rightarrow \bar{z}_\alpha$ and the commutators with the Poisson brackets $[ \ , \ ] \rightarrow -i \{ \ ,  \ \}$. By considering states of the form $\Psi(x)$, where $x_i =  \bar{z}_\alpha \sigma^i_{\alpha \beta} z_\beta$, that contain an equal number of $z$ and $\bar{z}$, one reproduces the ordinary QM in $\textbf{R}^3$. Allowing states with a fixed difference in the number of $z$ and $\bar{z}$ (for example of the form $\Psi(x) \cdot z_1^\kappa$) results into introducing monopole states of an arbitrary field strength $\mu = -\frac{\kappa}{2}$, more details can be found in \cite{mm}.

There are different lines of research currently being investigated. For example, the operator $\hat{S}_{05}=\lambda^{-1}\hat{r}$ plays a crucial role in the presented model. It can be used to define a dual velocity operator $\bar{V}_a = i \lambda^{-1}[\hat{r} , \hat{V}_a], a=1,...,4$, that can be used to enclose a larger algebraic structure with particular $\hat{r}$-dependent coefficients. Another, physically very appealing, option is to try to define a relativistic version of the model – a daunting task finally starting to seem possible using the presented underlying symmetries.

In \cite{mbh}, a Schwarzschild black hole with NC smeared singularity was analyzed, leading to multiple interesting results: infinite Hawking temperatures are avoided, a microscopic black hole does not evaporate completely, but leaves a Planck size remnant, possibly contributing to the observed dark matter density and also being able to generate ultra-high-energy rays. It might be possible to follow up on this research by analyzing a black hole with a nonzero magnetic charge. 

An important property of the presented model of NC QM is that the obtained results were always exact, there was no need for a perturbative approach. This signalizes a rich structure of underlying symmetries at work – some of which have been revealed now. From this, it seems obvious that achieving any further development of the model should be done by utilizing and generalizing the presented algebraic frameworks. 

\subsection*{Acknowledgment}
This research was partially supported by COST Action MP1405 and project VEGA 1/0985/16.

\end{document}